# Recombinant Sort: N-Dimensional Cartesian Spaced Algorithm Designed from Synergetic Combination of Hashing, Bucket, Counting and Radix Sort

Peeyush Kumar, Ayushe Gangal, Sunita Kumari*, Sunita Tiwari

Computer Science and Engineering, G. B. Pant Government Engineering College, Delhi 110020, India

Corresponding Author Email: sunitakumari@gbpec.edu.in

**ABSTRACT**

Sorting is an essential operation which is widely used and is fundamental to some very basic day to day utilities like searches, databases, social networks and much more. Optimizing this basic operation in terms of complexity as well as efficiency is cardinal. Optimization is achieved with respect to space and time complexities of the algorithm. In this paper, a novel left-field N-dimensional cartesian spaced sorting method is proposed by combining the best characteristics of bucket sort, counting sort and radix sort, in addition to employing hashing and dynamic programming for making the method more efficient. Comparison between the proposed sorting method and various existing sorting methods like bubble sort, insertion sort, selection sort, merge sort, heap sort, counting sort, bucket sort, etc., has also been performed. The time complexity of the proposed model is estimated to be linear i.e. $O(n)$ for the best, average and worst cases, which is better than every sorting algorithm introduced till date.

## 1. INTRODUCTION

Sorting is a process of arranging the given data into an ascending or a declining fashion on the basis of a linear relationship among the data elements [1]. Sorting may be performed on numbers, strings or records containing both numbers and strings like names, IDs, departments, etc., in alphabetical order, or in increasing or decreasing manner [2]. The exponential rise in the quantity of data available for use and being used, calls for more efficient and less time consuming sorting methods. Sorting algorithms are of two major types, namely, comparison and non-comparison sorting. Comparison sort involves sorting the data elements by doing repetitive comparisons and deciding which data element should come before or after which data element in the sorted array [3]. Comparison sort based sorting methods are bubble sort, insertion sort, quick sort, merge sort, shell sort, etc. Non-comparison sort does not compare the data elements for sorting them into an order. The non-comparison based sorting methods involve counting sort, bucket sort, radix sort, etc. Sorting algorithms can also be stable and unstable, in-place and out of place. In-place sorting algorithms are those which sort the given data without employing an additional data structure [4]. Out-of-place sorting algorithms require an additional or auxiliary data structure for sorting the given data elements [5]. Stable sort refers to the sorting technique in which two elements having equal values appear in the same order in the sorted array as they were, before the sorting was applied [6]. In the case of unstable sort, this order is not necessarily retained. Bubble sort, merge sort, counting sort and insertion sort are examples of stable sorting algorithms. While, quick sort, heap sort and selection sort are based on unstable sorting technique.

Each of these sorting algorithms have unique properties that add value to the specific function they are used to perform. Sorting algorithms are majorly distinguished on the basis of four properties, which are adaptability, stability, in-place/ not in-place, and online/ not-online, in addition to their basic methodology. An algorithm is adaptable in nature if its time complexity becomes almost $O(n)$ if the array is nearly sorted. An algorithm is said to display online property if it can process the input element by element, and doesn't require the whole array as input at the beginning. Bubble sort works by exchanging method and is an in-place, stable sorting algorithm, which makes $O(n^2)$ comparisons and swaps. It is not online and is adaptive in nature. Insertion sort works by insertion method and is also a stable, in-place sorting algorithm, which requires $O(n^2)$ comparisons and swaps. It is adaptive and online, in addition to having little over-head. Heap sort works by selection methodology and makes use of heap data structure. Both heap sort and quick sort are unstable in nature, and takes $O(n \log n)$ for comparisons and swaps. Heap sort is not-online, not-adaptive and is an in-place sorting algorithm, while quick sort is not-online, adaptive and an in-place sorting algorithm. Quick sort also has less over-head and works by partitioning. Bucket sort is a type of non-comparison distribution sort, which is not-online, out-of-place, non-adaptive and stable in nature. It has overheads of the buckets. Radix sort is a non-comparison integer sort, which is stable, not-online, adaptive and in-place in nature. In this paper, a novel sorting method is proposed by combining all the best characteristics of a few existing sorting algorithms. This novel method, called Recombinant Sort, combines the counting sort, bucket sort and radix sort, along with hashing and dynamic programming to elevate efficiency. This selective combination precedes the sum of the qualities of its parent algorithms, which brings out the essence of the idea behind this synergy. The proposed method has many unique and striking properties. It can work on numbers as well as strings, and can sort numbers containing decimals as well as non-decimal numbers together or apart.

Due to the application of hashing and dynamic programming, the traversal for fetching values is decreased tremendously and thus, the time complexity is also reduced. Comparison of various existing sorting algorithms is also conducted, on the basis of best, average and worst cases of time complexity, the ability to process decimals and strings, stability and on in-place or out-of-place technique.

This paper is divided into seven sections. Section 2 elaborates all the concepts used as pre-requisites for the proposed method. Section 3 delineates the concept, algorithm and the working of the proposed methodology of the recombinant sort. Proper description of algorithms, along with labelled diagrams are used to enhance the readers' understanding, and highlight the proposed novel approach in a lucid manner. Section 4 provides the proof of correctness of the proposed algorithm using loop invariant method. Section 5 contains the complexity analysis of the proposed algorithm and section 6 discusses the results obtained in a graphical and neatly tabulated manner. Section 7 discusses the conclusions and the future prospects of this algorithm and the domain.

## 2. CONCEPTS USED

### 2.1 Hashing

Hashing is a faster and more efficient method of insertion and retrieval of data. It works by employing a function called the hash function, which is used for generating new indices for the data elements. The hash function applies a uniform mathematical operation to all the data elements to allot them a place in the hash table. A hash table is a data structure that stores the values mapped by the hash function [7]. With hashing, the speed of retrieval or insertion can't be known but space-time trade off comes to picture. The speed can be checked by using a known amount of space for hashing, or the space used can be checked using a known speed for the process. Though usually, the speed of searching, insertion and deletion in hash tables is fast if collision of data does not occur, but it still heavily depends on the selection of the hash function. As hashing works by inducing randomness in the hash table and not order, it can't be considered to perform an admirable job for sorting the data alone. Hashing becomes extremely inefficient as the number of collisions increases, which causes the number of tuples in a bucket to increase, and ultimately leads the time complexity to become more linear $O(n)$. Hashing is used for a variety of applications, like, password verification, Rabin-Karp algorithm, compiler applications, message digest and in linking file name to path.

### 2.2 Bucket sort

Bucket sort works by distributing the data elements to be sorted in different buckets, which are then individually sorted using any other sorting technique or by recursive application of the bucket sort technique itself. The complexity of bucket sort depends on the number of buckets used, algorithm used for sorting each bucket and the uniformity in distribution of the data elements [8]. Once the elements are sorted into different buckets, the sorting of elements of the bucket becomes an independent task, and thus can also be carried out in parallel with other buckets to enhance performance. It can't be applied for string data type and requires a high degree of parallelism for achieving good performance [9]. Also, a bad distribution of elements in the buckets may very easily lead to extra work and degraded performance. Time complexity of bucket sort is $O(n+k)$ for best and average cases and $O(n^2)$ for the worst case. Bucket sort works best when the input data is of floating point type and is distributed uniformly over a range.

### 2.3 Counting sort

Counting sort is a small integer sorting technique, which works by counting data elements with distinct key values. Arithmetic is applied on these counts to determine the positions of the elements in the output. It is only suitable for data items in which the variation in the values of the elements do not precede the total number of elements to be sorted, as it has linear running time in total number of elements and difference between the maximum key and the minimum key values [10]. It is a stable sort and does not work by doing comparisons, thus is a non-comparison sort. Counting sort's time complexity is $O(n+k)$, where n is the size of the sorted array and k is the size of the helper array, which is needed when sorting non-primitive elements. Counting sort uses the values of the keys as indices, thus is only suitable for sorting small integers and can't be used to sort large datasets. As it only works for discrete values, it can't be used to sort strings and decimal values as array frequencies cannot be constructed. Counting sort has linear time complexity of $O(n+k)$ for the elements within the range of 1 to k, but turns to $O(n^2)$ for elements within the range of 1 to $n^2$ [11]. Counting sort is used when linear time complexity is needed and there are multiple entries of smaller magnitude integers.

### 2.4 Radix sort

Radix sort is a non-comparison sorting algorithm that works by considering the radix of the elements for distributing them into different buckets. The process of bucketing is repeated for each digit, with the previous ordering being preserved, again if the elements contain more than one significant digit [12]. Therefore, it is fast when the keys are small and the range of the array is less. Radix sort is known to be a close cousin of the counting sort. Though radix sort can work for integers, words, or any other dataset which can be lexicographically sorted, its flexibility is curbed as it depends on digits or letters to perform sorting. Separate codes need to be written for integers, floating type values and for strings. It is slower in comparison to merge sort and quicksort when the operations like insertion and deletion are not efficient enough and also has high space complexity [13]. The radix sort's constant k in $O(kn)$ is greater in comparison to any other sorting algorithm, and radix sort also consumes much greater space than quick sort, which is an in-place sorting algorithm. Radix sort is mostly used for sorting strings like stably sorting fixed-length words over fixed alphabets.

## 3. PROPOSED RECOMBINANT SORT ALGORITHM

The Recombination Sort is formulated from recombination of cardinal concepts of various sorting algorithms. The capability of radix sort to deal with each digit of the number separately, the concept of counting the number of occurrences of the elements in counting sort, the concept of bucketting from bucket sort and the concept of hashing a number to a multidimensional space are combined together to form a single sorting algorithm which outperforms its parent algorithms. As Radix sort is one of the parent algorithms, the recombinant sort needs to be rewritten for every different type of data. The Recombinant Sort consists of two parts, namely, the Hashing cycle and the Extraction cycle. For the purpose of simplicity, an array consisting of numbers between the of range 1 to 10, consisting of only one digit after decimal, is considered.

**3.1 Hashing cycle**

3.1.1 Mathematical rendition of hashing used in hashing cycle

For an *n*-digit decimal number $\Theta = n_1 n_2 n_3 \ldots n_{\lambda-1} n_\lambda . n_{\lambda+1} n_{\lambda+2} n_{\lambda+3} \ldots n_{n-1} n_n$ , $\forall \lambda \in Z$ , the hash function $H(\Lambda_\Theta)$, where $\Lambda_\Theta$ = set containing all digits of decimal number $\Theta$ in a systematic order from left to right : $(n_1, n_2, n_3, \ldots, n_{\lambda-1}, n_\lambda, n_{\lambda+1}, n_{\lambda+2}, n_{\lambda+3}, \ldots, n_{n-1}, n_n)$, can be defined as:

$$H(\Lambda_\Theta) = \{S[n_1][n_2][n_3]\ldots[n_{\lambda-1}][n_\lambda][n_{\lambda+1}][n_{\lambda+2}][n_{\lambda+3}]\ldots[n_{n-1}][n_n] + +\} \quad (1)$$

where, *S* is an *n*-dimensional cartesian space initialized by the hash function in the form of a hypercube to map an *n* digit number $\Theta$. The '++' sign donates an increment by 1. This increment by 1 is used in the hash function to tackle the problem of collision in hashing, thus the need for chaining list data structure is eliminated. Each axis of each dimension of *S* lies from [0, 9] and for an *n*-digit number $\Theta$, the shape of the space *S* initialized in the computer's memory is in the form of a hypercube (*n*-dimensional array) with each axis consisting of only 10 memory blocks, and can be expressed as:

$$shape(S) \equiv S[10][10][10]\ldots[10][10][10]\ldots[10][10] \quad (2)$$

The hash function defined in Eq. (1) maps a number $\Theta$ to an n-dimensional array *S*, defined in Eq. (2). The main goal in hashing is to minimize the time complexity [14] of the whole hashing operation. From Eq. 1, it can be stated that the hash function updates/increment (or maps the number $\Theta$ at) the index $\Lambda_\Theta$ of hypercube/array *S*. As the updation or deletion or fetching in an array has the time complexity of *O*(1) for each element [14], therefore the time complexity of hash function $H(\Lambda_\Theta)$ for each element is also *O*(1). Thus, the hash function maintains the minimum time complexity that can be maintained by a hash function and along with it, due the use of a hypercube/array data structure as hash table, the traversing through the table is also fast as well as continuous, and unlike counting sort, the large numbers can be sorted using space *S*.

3.1.2 Assumed pre-conditions

Only a single main precondition is required to instantiate the hashing cycle for the entire data consisting of *N* elements, which is, that each element should have the same number of digits, i.e, if elements does not have same number of digits then additional zeros are added to make up for the few digits in a way that it doesn't effect or change the quantity of the number. For example, if we have three numbers: [1.01, 2.1, 1], so in order to make these numbers have an equal number of digits, we add zeroes: [1.01, 2.10, 1.00]. This step is extremely easy and does not affect the efficiency of the algorithm in any magnitude. This step is also cardinal to keep track of the decimal's position. After this step, the unsorted array are given in Figure 1, defined as:

*arr* = [ 4.5, 0.3, 2.3, 8.8, 7, 9.2, 4.5, 4.3, 8, 3.2] can be written (after doing the preprocessing) as:

*arr* = [ 4.5, 0.3, 2.3, 8.8, 7.0, 9.2, 4.5, 4.3, 8.0, 3.2]

3.1.3 Dynamic programming used in $H(\Lambda_\Theta)$

As for an *n* digit decimal number, an initialized *n*-dimensional cartesian space will have a lot of unused space left after all the numbers have been mapped. But to increase the efficiency of the algorithm in order to retrieve the filled spaces (as in computer, in order to travel to or find the filled memory spaces in an array, one has to travel in a systematic pattern), a trick by maintaining two separate maps has been also performed. A more understood definition of two such maps has been given in the Hashing Cycle section.

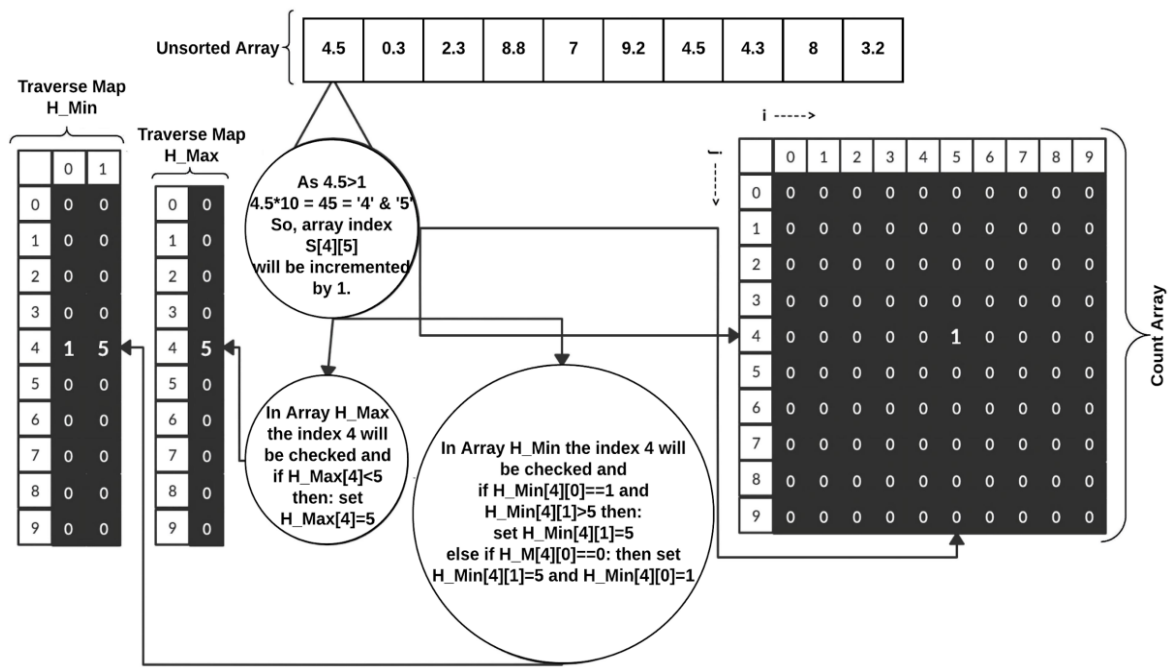

**Figure 1.** Hashing cycle

The steps of the hashing cycle are lucidly depicted in Figure 1 (the hashing function $H(\Lambda_\Theta)$ defined above is used for each element of the unsorted array *arr*). For sorting the type of data considered in the example, a 2D array of dimension 10x10 called the Count array, where the values will be mapped is considered, a traverse map H_Max of dimension 10x1 and a traverse map H_Min of dimension 10x2 are also taken. The two traverse maps are used to avoid unnecessary steps during the extraction period. The algorithm for hashing cycle designed for the example considered is as follows:

___

**HASHING CYCLE ALGORITHM**: The algorithm presented below uses two function: First, the numeric to string converter function, defined as: $F_{string}()$ and second, the string to numeric converter, defined as: $F_{Numeric}()$.
___

**Recombinant-hashing**(*arr*, size)**:**     //the unsorted array *arr*
    S[10][10];
    H_Max[10];
    H_Min[10][2];
    set digit_count_after_decimal ← 1
    **for** i = 0 to size **do**
        $t = F_{string}(arr[i] \times (10^{digit\_count\_after\_decimal}))$
        $S[F_{Numeric}(t[0])][F_{Numeric}(t[1])]$ ← *increment* by *1*
        **if**( $H\_Max[F_{Numeric}(t[0])] < F_{Numeric}(t[1])$) **then**
            set $H\_Max[F_{Numeric}(t[0])] \leftarrow F_{Numeric}(t[1])$
        **if** ($H\_Min[F_{Numeric}(t[0])][0] == 0$) **then**
            set $H\_Min[F_{Numeric}(t[0])][1] \leftarrow F_{Numeric}(t[1])$
            set $H\_Min[F_{Numeric}(t[0])][0] \leftarrow 1$
        **else if** ($H\_Min[F_{Numeric}(t[0])][0] \neq 0$ **and** $H\_Min[F_{Numeric}(t[0])][1] > F_{Numeric}(t[1])$) **then**
            set $H\_Min[F_{Numeric}(t[0])][1] \leftarrow F_{Numeric}(t[1])$
    **end for( i )**
**end func**
___

As depicted in Figure 1, the array *arr* (defined above) is fed to the hashing cycle for sorting and the space *S* of 10x10 is initialized along with a vector H_Max of shape 10 and a space H_Min of shape 10x2. The further steps are as follows:
1. The first element of the array is '4.5', so:
    a. First, it will be multiplied by $10^1$ (as count after decimal is 1): 4.5×10 = 45
    b. Second, the number 45 will be converted to string using $F_{String}(45)$ = t = '45'.
    c. Third, we will increment the value in the memory block at row t[0] = 4 and column t[1] = 5 (at array index (4, 5)).
    d. Fourth, in the traverse map H_Max, as H_Max [t[0]] < t[1], then H_Max[t[0]] will be set as t[1].
    e. Fifth, in traverse map H_Min, as H_Min [t[0]][0] = = 0, then H_Min[t[0]][1] will be set as t[1] and H_Min[$F_{Numeric}$(t[0])][0] will be set to 1.
    f. Lastly, the complete process will continue for each array element till we reach the end of the unsorted array and the rest of the complete steps are given in the example 1 in the supplementary section.

## 3.2 Extraction cycle

The end result of the hashing cycle is depicted in Figure 2. For the extraction of sorted arrays from the count array, the exaction cycle moves row by row, like done in raster scanning, for example, the cycle will visit all indices of row 0 and then all indices of row 1 and so on.

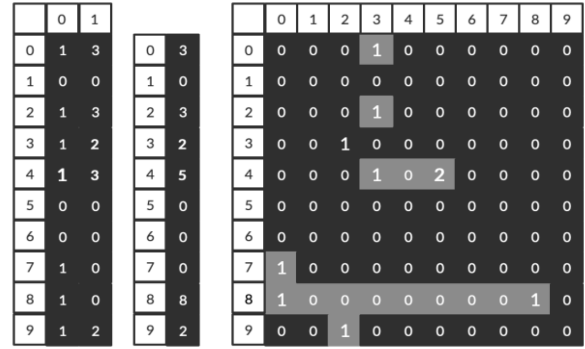

**Figure 2.** Extraction cycle

It is clear from Figure 2 that most of the memory spaces in the count array are not filled and thus, traversing these unused spaces will increase the time complexity of the algorithm. So, in order to minimize the time complexity and prevent wasteful traversal of these unused spaces, the traverse maps H_Min and H_Max are used. In the Traverse map H_Min, each row stores the lowest numeric value attained by the columns for that particular row in the count array and in the Traverse map H_Max, each row stores the highest numeric value attained by the columns for that particular row in the count array, for example, for row 4 in the count array the minimum column reached is 3 and the maximum column reached is 5, so the traverse map H_Min will store the value 3 and of H_Max will store the value 5, for row 4 of the count array. The column 0 of the traverse map H_Min will store whether the map for that particular row had been updated before or not. The algorithm for extraction cycle is as follows:

___
**EXTRACTION CYCLE ALGORITHM**: The algorithm presented below uses a function defined as: *Overwrite_arr(element, position, arr)*. This function overwrites the element '*element*' at position '*position*' of array '*arr*'. The importance of pre-conditions, defined above, can be seen in line 7. The $F_{Float}()$ function used below converts strings to float numbers and numeric to string converter function is also used and is defined as: $F_{string}()$. In line 7, the '+' sign is used to represent string concatenation.
___

**Recombinant-extraction**(S, H_Min, H_Max,*arr,* size)**:**
    set overwrite_pos_at ← 0
    **for** i = 0 to 9 **do**
        **for** j = H_Min[i][1] to H_Max[i]+1 **do**
            **if**(S[i][j]!=**Empty**) **then**
                **for** z = 0 to S[i][j] **do**
                    **Overwrite_arr**($F_{Float}(F_{string}$(i)+'.'+ $F_{string}$(j)), overwrite_pos_at, *arr*)
                    overwrite_pos_at ← **increment** by 1
                    **if**(overwrite_pos_at == size) **then**
                        **return** *arr*
                        **end func**

```
        end for( z )
     end for( j )
  end for( i )
end func
```
___________________________________________

At the end of the extraction cycle, the parts of the count array traversed are shown in Figure 2 and the sorted array obtained is shown below. The time complexity of the example taken in the prior section is found to be $O(n+17)$. Sorted array *arr* returned as:

[0.3, 2.3, 3.2, 4.3, 4.5, 4.5, 7.0, 8.0, 8.8, 9.2]

**Why Does Extraction Cycle Work?**

As it is known that in a computer's memory, in order to travel through an *n*-dimensional space one has to travel in a systematic pattern, performed using for loops. This systematic pattern traversal has a unique ability, whose advantage has been taken during the extraction cycle. A similar pattern can be observed when traversing a Binary Search Tree in an inorder traversal fashion. By traversing inorderly, a sorted form of the unsorted data used to build the binary search tree can be obtained. Due to the unique hash function $H(\Lambda_\theta)$ used to map an *n*-digit number $\theta$ to an *n*-dimensional cartesian space, a unique sorted outcome observed when performing inorder traversal in a binary search can also be observed when using Extraction Cycle (proposed above) to traverse through an *n*-dimensional cartesian space.

## 4. PROOF OF CORRECTNESS

Loop Invariant Induction [14] method has been used to prove the correctness of the proposed algorithm. It has been used to prove the correctness of both the Hashing Cycle and the Extraction Cycle. The correctness is for the recombinant sorting algorithm for *n* digit decimal numbers (the pseudo code for which is given in Supplementary Section).

***Notations***: The unsorted array is denoted by *arr[ ]* and *S* is used to denote the initialized *n*-dimensional space. The two arrays used to lower the extraction cost of the algorithm are denoted by *H_Min* and *H_Max*. It is assumed that *arr[ ]* contains *N* elements and the precondition stated above has been satisfied, and therefore, each element of *arr[ ]* has *n* digits. It is also assumed that the decimal point is placed after $\lambda^{th}, \forall \lambda \in Z$, digit for each element and thus, $\lambda$ digits lie before the decimal and ($n-\lambda$) digits lie after decimal for each element.

### 4.1 Hashing cycle

The predominant objective of this function is to use $H(\Lambda_\theta)$ so as to map all n elements of *arr[ ]* to an n-dimensional cartesian space, which is in the form of an array in the computer's memory. So, for every element in *arr[ ]*, there exists a place in the n-dimensional space. Therefore, the loop invariant $I_{Hash}$ for iteration $i^{th}$ can be defined as,

$I_{Hash} \equiv$ *At iteration i, the initialized n-dimensional empty cartesian space S should have $\leq$ i points mapped in it by using $H(\Lambda_\theta)$. Also, the arrays H_Min and H_Max should have $\leq 2i$ and $\leq i$ spaces mapped in it respectively.*

Or

$I_{Hash} \equiv$ *At iteration i, the cycle should successfully map all elements in arr[0: i] using $H(\Lambda_\theta)$, to an n-dimensional space S, initialized prior to the starting of loop.*

The three steps for Loop invariant proof are as follows:

(1) *Initialization*: Before the first iteration of the loop or at $i=0$ in the cycle, the invariant $I_{Hash}$ states that the initialized *n*-dimensional empty cartesian space should have $\leq 0$ points mapped in it by using $H(\Lambda_\theta)$. Also, the arrays *H_Min* and *H_Max* should have $\leq 2*0$ and $\leq 0$ spaces mapped in it, respectively. As 0 points have been mapped in space *S* at $i=0$, therefore the space remains vacant. Also, 0 spaces in arrays *H_Min* and *H_Max* have been mapped, therefore they also remain unoccupied. As the space *S and* arrays *H_Min* and *H_Max* were already set to be vacant, the invariant condition stands corrected.

(2) *Maintenance*: Assume that the loop invariant stands corrected at the start of iteration $i=j$ in the cycle. Then it must be that the initialized n-dimensional empty cartesian space S should have $\leq j$ points mapped in it by using $H(\Lambda_\theta)$. Also, the arrays *H_Min* and *H_Max* should have$\leq 2*j$ and $\leq j$ spaces mapped in it respectively. In the body of the loop at iteration *j*, *arr[j]* is mapped to cartesian space *S,* and if the defined condition holds *True,* then the required values are mapped in arrays *H_Min* and *H_Max*. Thus, at the start of the iteration $i = j+1$, the initialized n-dimensional empty cartesian space S will have $\leq j + 1$ points mapped in it by using $H(\Lambda_\theta)$. Also, the arrays *H_Min* and *H_Max* will have $\leq 2*(j + 1)$ and $\leq j + 1$ spaces mapped in it respectively, which needed to be proved.

(3) *Termination*: When the ***for***-loop terminates at $i = N$, the initialized n-dimensional empty cartesian space S has $\leq N$ points mapped in it by using $H(\Lambda_\theta)$. Also, the arrays *H_Min* and *H_Max* has $\leq 2*N$ and $\leq N$ spaces mapped in it respectively. As *arr[ ]* has *N* elements, therefore all elements have been mapped to space *S*, which is also the desired output.

As all three steps of the loop invariant hold true, therefore the algorithm for the hash cycle is correct.

### 4.2 Extraction cycle

It is a known fact that a human can traverse an *n*-dimensional space in either a linear or a nonlinear fashion, but computers can only traverse such spaces in a linear fashion. This linear traversal, as also mentioned before, yields an advantage. For instance, in order to traverse through a 2-dimensional space of size 10×10, a ***for*** or ***while***-loop is needed, and upon discerning those loops closely, one would notice that they are intrinsically counting from 0 to 100 (and undeniably counting is sorted). Thus, the extraction cycle traverses the *n*-dimensional space in a fashion that it encounters the mapped elements in a sorted manner, due to the intrinsic nature of loop traversal. $n+1$ *number of **for***-loops are required for traversing an *n*-dimensional array, as well as for extracting the numbers mapped. In order to define $n+1$ ***for***-loops, $n+1$ number of iterators will be needed, and can be defined as: $(i_1, i_2, i_3, \ldots, i_{n-1}, i_n, i_{n+1})$. Another variable *overwrite_at_pos* is defined to keep track of how many occupied spaces in *S* have been detected and tells where to overwrite the original unsorted array. The loop invariant $I_{Extract}$ for every $i^{th}$ iteration can be defined as,

$I_{Extract} \equiv$ *At iteration overwrite_at_pos = j and iterators $i_1, i_2, i_3, \ldots, i_{n-1}, i_n$ and $i_{n+1}$ having any value such that the mentioned **if**-condition is satisfied, the element $E_{j+1}$ detected* at $S[i_1][i_2][i_3]\ldots[i_{n-1}][i_{n+1}]$ can be represented as:

$$E_1, E_2, E_3, \ldots, E_j \leq E_{j+1} \leq E_{j+2}, E_{j+3}, E_{j+4}, \ldots, E_N \quad (3)$$

And the overwritten sub-array *arr[0:j]* should be sorted or the sub-array *arr[j:N]* should be unsorted or unchanged.

The three steps for Loop invariant proof are as follows:
(1) *Initialization*: Before the first iteration of the loop or at *overwrite_at_pos=j=0* and iterators $i_1, i_2, i_3, \ldots, i_{n-1}, i_n, i_{n+1}$ having any value such that the mentioned *if*-condition is satisfied, the element $E_1$ detected at $S[i_1][i_2][i_3]\ldots[i_{n-1}][i_{n+1}]$ can be represented as:

$$E_1 \leq E_2, E_3, E_4, \ldots, E_N \quad (4)$$

And the overwritten sub-array *arr[0:0]* should be sorted or the sub-array *arr[0:N]* should be unsorted or unchanged. As the size of the overwritten sub-array *arr[0:0]* is zero, or the overwritten sub-array *arr[0:0]* is completely empty, therefore, it is sorted. Also, as the sub-array *arr[0:N]* was already unsorted or unchanged, the invariant condition stands corrected.

(2) *Maintenance:* Assume that the loop invariant stands corrected at the start of iteration *overwrite_at_pos=j=z* and and iterators $i_1, i_2, i_3, \ldots, i_{n-1}, i_n, i_{n+1}$ having any value such that the mentioned *if*-condition is satisfied in the cycle. Then it must be that the element $E_{z+1}$ detected at $S[i_1][i_2][i_3]\ldots[i_{n-1}][i_{n+1}]$ can be represented as:

$$E_1, E_2, E_3, \ldots, E_{z-1} \leq E_{z+1} \leq E_{z+2}, E_{z+3}, E_{z+4}, \ldots, E_N \quad (5)$$

And the overwritten sub-array *arr[0:z]* should be sorted or the sub-array *arr[z:N]* should be unsorted or unchanged. In the body of the loop at iteration *overwrite_at_pos=j=z,* the extracted element is overwritten at index *z* of unsorted array *arr[]*, leaving sub-array *arr[0:z]* sorted or sub-array *arr[z:N]* unsorted or unchanged. Thus, at the start of iteration *overwrite_at_pos=j=z+1* and iterators $i_1, i_2, i_3, \ldots, i_{n-1}, i_n, i_{n+1}$ having any value such that the mentioned *if*-condition is satisfied in the cycle, the element $E_{z+2}$ detected at $S[i_1][i_2][i_3]\ldots[i_{n-1}][i_{n+1}]$ will be represented as:

$$E_1, E_2, E_3, \ldots, E_{z+1} \leq E_{z+2} \leq E_{z+3}, E_{z+4}, E_{z+5}, \ldots, E_N \quad (6)$$

And the overwritten sub-array *arr[0:z+1]* will be sorted or the sub-array *arr[z+1:N]* will be unsorted or unchanged, which needed to be proved.

(3) *Termination*: When the *for*-loop terminates at *overwrite_at_pos=j=N* and iterators $i_1, i_2, i_3, \ldots, i_{n-1}, i_n, i_{n+1}$ having any value such that the mentioned *if*-condition is satisfied in the cycle, then the element $E_{N+1}$ detected at $S[i_1][i_2][i_3]\ldots[i_{n-1}][i_{n+1}]$ is represented as:

$$E_1, E_2, E_3, E_4, \ldots, E_N \leq E_{N+1} \quad (7)$$

As $E_{N+1}$ does not exist, therefore, all *N* elements have been detected and overwritten. Also, the overwritten sub-array *arr[0:N]* will be sorted or the sub-array *arr[N:N]* will have size zero or contain zero elements, which is also the desired output.

As all three steps of the loop invariant hold true, therefore the algorithm for the extraction cycle is correct. Also, as both the hashing cycle and extraction cycle are correct, it renders the proposed Recombinant Sort algorithm correct. The correctness of the algorithm can also be verified from the example 1 described in the supplementary section.

## 5. COMPLEXITY ANALYSIS

**Best case:** The best case takes place when the extraction cost *k* (= total number of memory block traversed in hypercube S during the whole extraction cycle process) will be of the form, $k<<<n$ and thus, the time complexity $O(n+k)$ will be $O(n)$. The possible scenarios of best cases (where cost k is minimum) are as follows:
i. If all elements of the unsorted array lie on the same horizontal axis (after mapping) of the hypercube space S.
ii. If all elements of the unsorted array lie on the same vertical axis (after mapping) of the hypercube space S.
iii. If all elements of the unsorted array lie inside the same memory block (after mapping) of the hypercube space S.

**Average case:** The Average case takes place when the extraction cost *k* will be of the form, $k<=n$. The possible cases for the average time complexity will be as follows:

i. If $k<n$, then the time complexity $O(n+k)$ will be $\equiv O(n)$ after taking the upper bound n.
ii. If k = n, then the time complexity $O(n+k)$ will be $O(2n) \equiv O(n)$.

Thus the average time complexity in both possible cases is $O(n)$.

**Worst case:** The Worst case takes place when one of the two possible cases defined below happens:
i. *First*: When during the extraction cycle, the whole count array needs to be traversed. Thus, making the extraction cost $k=10^b$, where *b* = the maximum number of digits an element has in our dataset. But for reaching worst case the count array has to be filled completely, thus, at least $10^b (n=10^b)$ elements have to be there in the dataset. Therefore, the time complexity $O(n+k)$ will be:

$$O(n+10^b) \quad (8)$$

But $n=10^b$, so

$$O(n+n)=O(2n)=O(n) \quad (9)$$

ii. Second: When the start and end memory block of each axis of hypercube space *S* is occupied and the rest of the memory block in between them are empty. Thus, in this case the whole space *S* needs to be traversed, which makes the extraction cost $k=10^d$, where *d* = dimensions of hypercube space *S* or the maximum number of digits an element has in our dataset. And the time complexity will be:

$$O(n+10^d) \quad (10)$$

But for this case to be valid, the total number of elements to be sorted should be =10d. Thus, it can be stated that *n*=10*d* and the Eq. (10) can be written as:

$$O(10d + 10^d) \quad (11)$$

$$O\left(10d\left(1 + \frac{10^{(d-1)}}{d}\right)\right) \quad (12)$$

As *n* =10*d*,

$$O\left(n\left(1 + \frac{10^{(d-1)}}{d}\right)\right) \quad (13)$$

Which can be further simplified as,

$$O(nC) \quad (14)$$

where, $C = \left(1 + \frac{10^{(d-1)}}{d}\right)$.

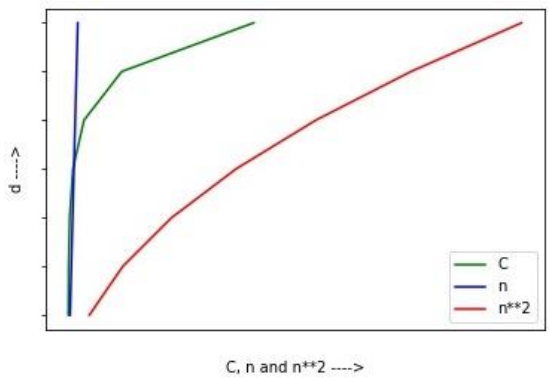

**Figure 3.** Relationship between constant *C*, *n*, *n²* and *d*

On the basis of experiment, whose results are shown in Figure 3, (by putting different values of d) it was observed that:

$$C < n^2 \quad (15)$$

Thus, from Eq. (15) we can state that C is a constant that will never make the time complexity nonlinear and the complexity given in Eq. (14) can be written as *O(n)*.

Thus, the time complexity will always be linear.

## 6. RESULTS AND DISCUSSIONS

Table 1 shows the time taken by the system to execute recombinant sort using Python on Mac OS. The sorting method is executed in Python, C++ and Java language on Mac OS, Windows OS and Linux OS. The system had 3.1 GHz Intel Core i5 processor and 8 GB 2133 MHz LPDDR3 RAM.

The testing data is generated from a random generator function available in python's numpy library. The number of elements taken for the execution of recombinant sort ranges from 10 to 10000, increasing in powers of 10. Time is calculated for five major cases, namely, for data between the range of 1 to 10 having no digits after the decimal, having a single digit after the decimal and having two digits after the decimal, and for data between the range of 1 to 100 having no digits after the decimal and having a single digit after the decimal. The time taken for a specific number of elements for all the cases are of comparable order, as can be seen from Table 1. The results obtained by running the algorithm in different languages on different operating systems platforms are shown in tabular format (Tables 4-11), along with their graphical representation (Figure 5), are given in the supplementary section. These languages (Python, C++ and Java) and operating systems (Window, Mac and Linux) have been chosen specifically as they are very widely used.

Figure 4 depicts the results obtained in Table 1 in a graphical manner. The graph shows the time taken to execute recombinant sort (in milliseconds) for all the five enlisted cases. The graph depicts linear characteristics of the proposed sorting algorithm and it can also be concluded from the graph that the *count_after_decimal* variable hardly affects the time complexity. These same observations can also be made while observing the graphs (for Python, C++ and Java languages and Mac OS, Windows OS and Linux OS) given in the supplementary section.

Table 2 gives a complete comparison between various existing and known sorting algorithms and the proposed Recombinant sort technique on the basis of various cardinal factors like best case time complexity, average case time complexity, worst case time complexity, stability, in-place or out-of-place sorting and the ability to process strings and floating point numbers.

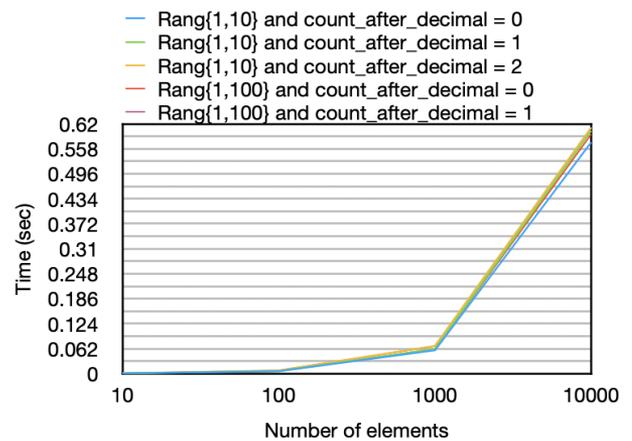

**Figure 4.** Relationship between the number of elements and the time taken by Recombinant Sort to sort elements using Python on Mac OS

**Table 1.** The time taken (in sec) by the system to execute recombinant sort using python on Mac OS

| No. of elements | TFD (1,10) & cd=0 | TFD (1,10) & cd=1 | TFD (1,10) & cd=2 | TFD (1,100) & cd=0 | TFD (1,100) & cd=1) |
|---|---|---|---|---|---|
| 10 | 0.00066 | 0.00078 | 0.00091 | 0.00079 | 0.00091 |
| 100 | 0.00581 | 0.00649 | 0.00792 | 0.00641 | 0.007.91 |
| 1000 | 0.5790 | 0.06124 | 0.06889 | 0.06112 | 0.06821 |
| 10000 | 0.57410 | 0.60279 | 0.61084 | 0.59412 | 0.61038 |

Note: The expression TFD(a,b) & cd = c stands for: Time For sorting Data ranging between a to b, and count after decimal = c respectively.

**Table 2.** Comparison with other sorting algorithms

| Sorting algorithm | Best TC | Average TC | Worst TC | stable sort | PD | PS | IP |
|---|---|---|---|---|---|---|---|
| Bubble sort [15] | $O(n)$ (waas) | $O(n^2)$ | $O(n^2)$ | Yes | Yes | No | Yes |
| Selection Sort [16] | $O(n^2)$ | $O(n^2)$ | $O(n^2)$ | No | Yes | No | Yes |
| Insertion Sort [17] | $O(n^2)$ (waas) | $O(n^2)$ | $O(n^2)$ | Yes | Yes | No | Yes |
| Merge Sort [18] | $O(nlogn)$ | $O(nlogn)$ | $O(nlogn)$ | Yes | Yes | No | No |
| Quick Sort [19] | $O(nlogn)$ | $O(nlogn)$ | $O(n^2)$ | No | Yes | No | Yes |
| Bucket Sort [9] | $O(n+c)$ | $O(n+c)$ | $O(n^2)$ | Yes | Yes | No | No |
| Radix Sort [12] | $O(kn)$ (k∈Z) | $O(kn)$ (k∈Z) | $O(kn)$ (k∈Z) | Yes | No | Yes | No |
| Heap Sort [20] | $O(nlogn)$ | $O(nlogn)$ | $O(nlogn)$ | No | Yes | No | Yes |
| Tim Sort [21] | $O(n)$ (waas) | $O(nlogn)$ | $O(nlogn)$ | Yes | Yes | No | No |
| Shell Sort [22] | $O(n)$ (waas) | $O(n^2)$ | $O(n^2)$ | No | Yes | No | Yes |
| Counting Sort [11] | $O(n)$ | $O(n+c)$ | $O(n^2)$ | Yes | No | No | No |
| **Recombinant Sort** | $O(n)$ | $O(n)$ | $O(n)$ | No | Yes | Yes | No |

Note: waas stands for: When array is already sorted; TC stands for Time Complexity; PD: Can Sort or Process Decimals; PS: Can Sort or Process Strings; IP: Inplace Sort; Z: integer.

**Table 3.** Dimensions of elements required for sorting different types of data

| S. No. | Type of Data | Dimensions of Count Array | Dimensions of Traverse Map H_Min | Dimensions of Traverse Map H_Max |
|---|---|---|---|---|
| 1. | D(1,10) & cd = 0 | 10 | 2 | 1 |
| 2. | D(1,10) & cd = 1 | 10x10 | 10x2 | 10x1 |
| 3. | D(1,10) & cd = 2 | 10x10x10 | 10x22 | 10x11 |
| 4. | D(1,100) & cd = 0 | 10x10 | 10x2 | 10x1 |
| 5. | D(1,100) & cd = 1 | 10x10x10 | 10x22 | 10x11 |

Note: The expression D(a,b) & cd = c stands for: Data ranging between a to b, and count after decimal = c respectively.

From Table 2, it is observed that merge sort and heap sort have the consistent time complexity of $O(nlogn)$ for the best, average and worst case scenarios, but none of these sorting methods can be used to sort elements of string data type. Quicksort also has $O(nlogn)$ time complexity for the best and average cases, but resorts to being $O(n^2)$ for the worst case, i.e. when the array is already sorted in any order or when the array contains all identical elements. Tim sort has the time complexity $O(nlogn)$ for the worst and average cases and $O(n)$ for the best case (given that the array is already sorted). Unlike the sorting algorithms listed above, the proposed recombinant sort has consistent performance of $O(n)$ for the best case, average case and worst case scenarios. In addition to this, recombinant sort can also be used to sort elements of string data type and floating type. Therefore, it can be observed that the proposed recombinant sort performs best among all the listed sorting algorithms.

Table 3 specifies the dimensions of the elements constituting the recombinant sort, that is, the count array, and the H_Min and H_Max traverse maps, for sorting data elements that belong to the data specified in the five cases enlisted previously.

This table depicts a pattern that can be followed to deal with different types of data (not mentioned in the table) using Recombinant Sort.

## 7. CONCLUSION AND FUTURE WORK

The proposed Recombinant Sort is a dynamic sorting technique which can be modified as per the needs of the user and is designed to achieve utmost efficiency for sorting data of varied types and ranges. The time complexity of the proposed Recombinant Sort is estimated to be $O(n+k)$ for best, average and worst cases. The $k$ in $O(n+k)$ will become $n$ in the worse case scenario, but in no circumstance will n's order approaches two, i.e, $k$ will never approach $n^2$, thus, the complexity will never be $O(n^2)$. The extraction cost $k$, will always be very less than or equal to $n$, thus, the final time complexity will always be $O(n)$. Also, the extraction cost $k$ of the proposed Recombinant Sort came out to be much smaller than the extraction cost of any other linear sorting algorithms. The graph plotted between the number of elements and the time taken by recombinant sort to sort those elements depicts a linear characteristic.

All major highlighted demerits of the parent algorithms of the Recombinant Sort, i.e., counting sort, radix sort and bucket sort, are surmounted by Recombinant Sort. Recombinant Sort can process strings as well as numbers, and can also process both floating point and integer type numbers together. Though, with the increase in the number of digits in elements to be sorted, the dimensions of the count array will increase, and the complexity of the working of the algorithm will also increase. But an important thing to note here is that, in the physical world, we don't usually deal with numbers containing more than 10 digits, be it, marks obtained or the net salaries. By testing the algorithm on all possible types of data, it has been empirically proved that the proposed algorithm is correct, complete and terminates at the end. Thus, Recombinant Sort is a viable option from the user's perspective. In order to accredit fair competition, an open source library named Recombinant Sort has been released on github.

In the future, the proposed Recombinant Sort can be enhanced by sorting integer, string and floating type elements without rewriting the entire program for these specific needs. Another noteworthy addition to the current proposed algorithm can be made post-availability of advanced literature on N-dimensional space or hypercubes.

## SUPPLEMENTARY SECTION

---

**HASHING CYCLE ALGORITHM FOR *n* DIGIT NUMBER**: The algorithm presented below uses two function: First, the numeric to string converter function, defined as: $F_{string}()$ and second, the string to numeric converter, defined as: $F_{Numeric}()$.

**NOTE**: In day to day life we usually deal with 4-5 digit numbers.

---

| | | |
|---|---|---|
| | **Recombinant-hashing**(arr, size, $\lambda$) | // the unsorted array *arr* |
| 1. | S[10][10][10]..[10][10]; | // *n* dimensional count array S is initialized |
| 2. | H_Max[10][10][10]..[10][10]; initialized | // *(n-1)* dimensional traverse map H_Max is |
| 3. | H_Min[10][222...222]; there | // traverse map H_Min is initialized. *(n-1)* 2's are |
| 4. | **set** digit_count_after_decimal ← $\lambda$ | |
| 5. | **for** i = 0 to size **do** | |
| 6. | $t = F_{string}(arr[i] \times (10^{count\_after\_decimal}))$ | //converts number to string |
| 7. | $S[F_{Numeric}(t[0])][F_{Numeric}(t[1])]\ldots[F_{Numeric}(t[n-1])]$ ← ***increment*** *by 1* | |
| 8. | **if** ( $H\_Max[F_{Numeric}(t[0])][0] < F_{Numeric}(t[1])$) **then** | // checking H_Max |
| 9. | **set** $H\_Max[F_{Numeric}(t[0])][0] \leftarrow F_{Numeric}(t[1])$ | |
| 10. | **if** ($H\_Min[F_{Numeric}(t[0])][0] == 0$) **then** | // if H_Min traverse map had been updated before |
| 11. | **set** $H\_Min[F_{Numeric}(t[0])][1] \leftarrow F_{Numeric}(t[1])$ | |
| 12. | **set** $H\_Min[F_{Numeric}(t[0])][0] \leftarrow 1$ | // marking that the H_Min is updated |
| 13. | **else if** ($H\_Min[F_{Numeric}(t[0])][0] \neq 0$ and $H\_Min[F_{Numeric}(t[0])][1] > F_{Numeric}(t[1])$) **then** | |
| 14. | **set** $H\_Min[F_{Numeric}(t[0])][1] \leftarrow F_{Numeric}(t[1])$ | |
| 15. | . | |
| 16. | . | |
| 17. | . | |
| 18. | **if**( $H\_Max[F_{Numeric}(t[0])][222..221] < F_{Numeric}(t[n-1])$) **then** | // checking H_Max |
| 19. | **set** $H\_Max[F_{Numeric}(t[0])][10][10]...[10][10] \leftarrow F_{Numeric}(t[n-1])$ | |
| 20. | **if** ($H\_Min[F_{Numeric}(t[n-1])][222..220] == 0$) **then** | // if H_Min traverse map had been updated before |
| 21. | **set** $H\_Min[F_{Numeric}(t[n-1])][222..221] \leftarrow F_{Numeric}(t[n-1])$ | |
| 22. | **set** $H\_Min[F_{Numeric}(t[n-1])][222..220] \leftarrow 1$ | // marking that the H_Mi is updated |
| 23. | **else if** ($H\_Min[F_{Numeric}(t[n-1])][222..221] \neq 0$ **and** $H\_Min[F_{Numeric}(t[n-2])][222...221] > F_{Numeric}(t[n-1])$) **then** | |
| 24. | **set** $H\_Min[F_{Numeric}(t[n-2])][222..221] \leftarrow F_{Numeric}(t[n-1])$ | |
| 25. | **end for**( j ) | |
| 26. | **end func** | |

---

**EXTRACTION CYCLE ALGORITHM FOR *n* DIGIT NUMBER**: The algorithm presented below uses a function defined as: *Overwrite_arr(element, position, arr)*. This function overwrites the element '*element*' at position '*position*' of array '*arr*'. The $F_{Float}()$ function used below converts strings to float numbers and numeric to string converter function is also used and is defined as: $F_{string}()$.

---

    **Recombinant-extraction**(S, H_Min, H_Max, arr, size)

| | | |
|---|---|---|
| 1. | **set** overwrite_pos_at ← 0 | |
| 2. | **for** $i_1 = 0$ to 9 **do** | |
| 3. | **for** $i_2$ = H_Min[$i_1$][1] to H_Max[$i_1$][0]..[0]+1 **do** | // maps H_Min and H_Max are tallied |
| 4. | **for** $i_3$ = H_Min[i][2] to H_Max[$i_1$][0]..[1]+1 **do** | // maps H_Min and H_Max are tallied |
| 5. | . | |
| 6. | . | |
| 7. | . | |
| 8. | **for** $i_n$ = H_Min[$i_1$][22..21] to H_Max[$i_1$][10]..[10]+1 **do** | // maps H_Min and H_Max are tallied |
| 9. | **if**(S[i][j]!=**Empty**) **then** | |
| 10. | **for** z = 0 to S[i][j] **do** | // generates the hashed data |
| 11. | **Overwrite_arr**(extracted_element, overwrite_pos_at, arr) | |
| 12. | overwrite_pos_at ← **increment** by 1 | |
| 13. | **if**(overwrite_pos_at == size) **then** | |
| 14. | **return** arr | |
| 15. | **end func** | |
| 16. | **end for**( $i_{1n}$ ) | |
| 17. | **end for**( $i_{n-1}$ ) | |
| 18. | . | |
| 19. | . | |
| 20. | . | |
| 21. | **end for**( $i_1$ ) | |
| 22. | **end func** | |

**Note:** The '*extracted_element*' defined above in line 11 can be represented as:

$$extracted\_element = i_1 i_2 i_3 .... i_{\lambda-1} i_\lambda . i_{\lambda+1} i_{\lambda+2} i_{\lambda+3} ... i_{n-1} i_n \qquad (16)$$

**RESULTS OF EXECUTION OF RECOMBINANT SORT USING DIFFERENT LANGUAGES ON DIFFERENT OPERATING SYSTEMS SHOWN USING TABULAR AS WELL AS GRAPHICAL METHOD**

**Table 4.** The time taken (in sec) by the system to execute recombinant sort written in Python on Windows OS

| No. of elements | TFD(1,10) & cd=0 | TFD(1,10) & cd=1 | TFD(1,10) & cd=2 | TFD(1,100) & cd=0 | TFD(1,100) & cd=1) |
|---|---|---|---|---|---|
| 10 | 0.00062 | 0.00071 | 0.00091 | 0.00085 | 0.0009 |
| 100 | 0.0049 | 0.00641 | 0.00797 | 0.00644 | 0.00783 |
| 1000 | 0.0572 | 0.06119 | 0.06882 | 0.06123 | 0.07001 |
| 10000 | 0.5738 | 0.60282 | 0.61071 | 0.59415 | 0.61042 |

Note: The expression TFD(a,b) & cd = c stands for: Time For sorting Data ranging between a to b, and count after decimal = c.

**Table 5.** The time taken (in sec) by the system to execute recombinant sort written in Python on Linux OS

| No. of elements | TFD(1,10) & cd=0 | TFD(1,10) & cd=1 | TFD(1,10) & cd=2 | TFD(1,100) & cd=0 | TFD(1,100) & cd=1) |
|---|---|---|---|---|---|
| 10 | 0.00052 | 0.00063 | 0.00089 | 0.00083 | 0.00094 |
| 100 | 0.0041 | 0.00641 | 0.00791 | 0.0065 | 0.00777 |
| 1000 | 0.0569 | 0.06123 | 0.06872 | 0.0612 | 0.0702 |
| 10000 | 0.5681 | 0.60281 | 0.60039 | 0.5835 | 0.61061 |

Note: The expression TFD(a,b) & cd = c stands for: Time For sorting Data ranging between a to b, and count after decimal = c.

**Table 6.** The time taken (in sec) by the system to execute recombinant sort written in Java on Windows OS

| No. of elements | TFD (1,10) & cd=0 | TFD (1,10) & cd=1 | TFD (1,10) & cd=2 | TFD (1,100) & cd=0 | TFD (1,100) & cd=1 |
|---|---|---|---|---|---|
| 10 | 0.00059 | 0.00069 | 0.00089 | 0.00085 | 0.00092 |
| 100 | 0.0044 | 0.00641 | 0.00791 | 0.00644 | 0.00782 |
| 1000 | 0.0572 | 0.06127 | 0.06822 | 0.06123 | 0.0701 |
| 10000 | 0.5682 | 0.60281 | 0.61066 | 0.57354 | 0.61039 |

Note: The expression TFD(a,b) & cd = c stands for: Time For sorting Data ranging between a to b, and count after decimal = c.

**Table 7.** The time taken (in sec) by the system to execute recombinant sort written in Java on Mac OS

| No. of elements | TFD (1,10) & cd=0 | TFD (1,10) & cd=1 | TFD (1,10) & cd=2 | TFD (1,100) & cd=0 | TFD (1,100) & cd=1 |
|---|---|---|---|---|---|
| 10 | 0.00059 | 0.00071 | 0.00091 | 0.00085 | 0.00091 |
| 100 | 0.0046 | 0.00644 | 0.00792 | 0.00644 | 0.00785 |
| 1000 | 0.0574 | 0.06122 | 0.06885 | 0.06123 | 0.07011 |
| 10000 | 0.5732 | 0.60285 | 0.61069 | 0.59415 | 0.61039 |

Note: The expression TFD(a,b) & cd = c stands for: Time For sorting Data ranging between a to b, and count after decimal = c.

**Table 8.** The time taken (in sec) by the system to execute recombinant sort written in Java on Linux OS

| No. of elements | TFD (1,10) & cd=0 | TFD (1,10) & cd=1 | TFD (1,10) & cd=2 | TFD (1,100) & cd=0 | TFD (1,100) & cd=1 |
|---|---|---|---|---|---|
| 10 | 0.00055 | 0.0007 | 0.00089 | 0.00081 | 0.00093 |
| 100 | 0.0047 | 0.00642 | 0.00791 | 0.00642 | 0.00783 |
| 1000 | 0.0569 | 0.06122 | 0.06123 | 0.06885 | 0.07013 |
| 10000 | 0.573 | 0.61068 | 0.60285 | 0.59339 | 0.6104 |

Note: The expression TFD(a,b) & cd = c stands for: Time For sorting Data ranging between a to b, and count after decimal = c.

**Table 9.** The time taken (in sec) by the system to execute recombinant sort written in C++ on Windows OS

| No. of elements | TFD (1,10) & cd=0 | TFD (1,10) & cd=1 | TFD (1,10) & cd=2 | TFD (1,100) & cd=0 | TFD (1,100) & cd=1 |
|---|---|---|---|---|---|
| 10 | 0.00058 | 0.00071 | 0.00091 | 0.00083 | 0.00091 |
| 100 | 0.0045 | 0.00644 | 0.00792 | 0.00641 | 0.00782 |
| 1000 | 0.0575 | 0.06122 | 0.06885 | 0.06125 | 0.07021 |
| 10000 | 0.5731 | 0.59415 | 0.61069 | 0.60115 | 0.6104 |

Note: The expression TFD(a,b) & cd = c stands for: Time For sorting Data ranging between a to b, and count after decimal = c.

**Table 10.** The time taken (in sec) by the system to execute recombinant sort written in C++ on Mac OS

| No. of elements | TFD (1,10) & cd=0 | TFD (1,10) & cd=1 | TFD (1,10) & cd=2 | TFD (1,100) & cd=0 | TFD (1,100) & cd=1 |
|---|---|---|---|---|---|
| 10 | 0.0005 | 0.00065 | 0.0009 | 0.00085 | 0.00094 |
| 100 | 0.0044 | 0.00641 | 0.00791 | 0.00644 | 0.00782 |
| 1000 | 0.0572 | 0.06127 | 0.06882 | 0.06123 | 0.0701 |
| 10000 | 0.5682 | 0.5682 | 0.61039 | 0.5835 | 0.61066 |

Note: The expression TFD(a,b) & cd = c stands for: Time For sorting Data ranging between a to b, and count after decimal = c.

**Table 11.** The time taken (in sec) by the system to execute recombinant sort written in C++ on Linux OS

| No. of elements | TFD (1,10) & cd=0 | TFD (1,10) & cd=1 | TFD (1,10) & cd=2 | TFD (1,100) & cd=0 | TFD (1,100) & cd=1 |
|---|---|---|---|---|---|
| 10 | 0.00049 | 0.00061 | 0.00084 | 0.0008 | 0.00091 |
| 100 | 0.0036 | 0.0064 | 0.0079 | 0.0065 | 0.0077 |
| 1000 | 0.057 | 0.0612 | 0.0687 | 0.0612 | 0.0702 |
| 10000 | 0.5679 | 0.6028 | 0.6003 | 0.5835 | 0.6106 |

Note: The expression TFD(a,b) & cd = c stands for: Time For sorting Data ranging between a to b, and count after decimal = c.

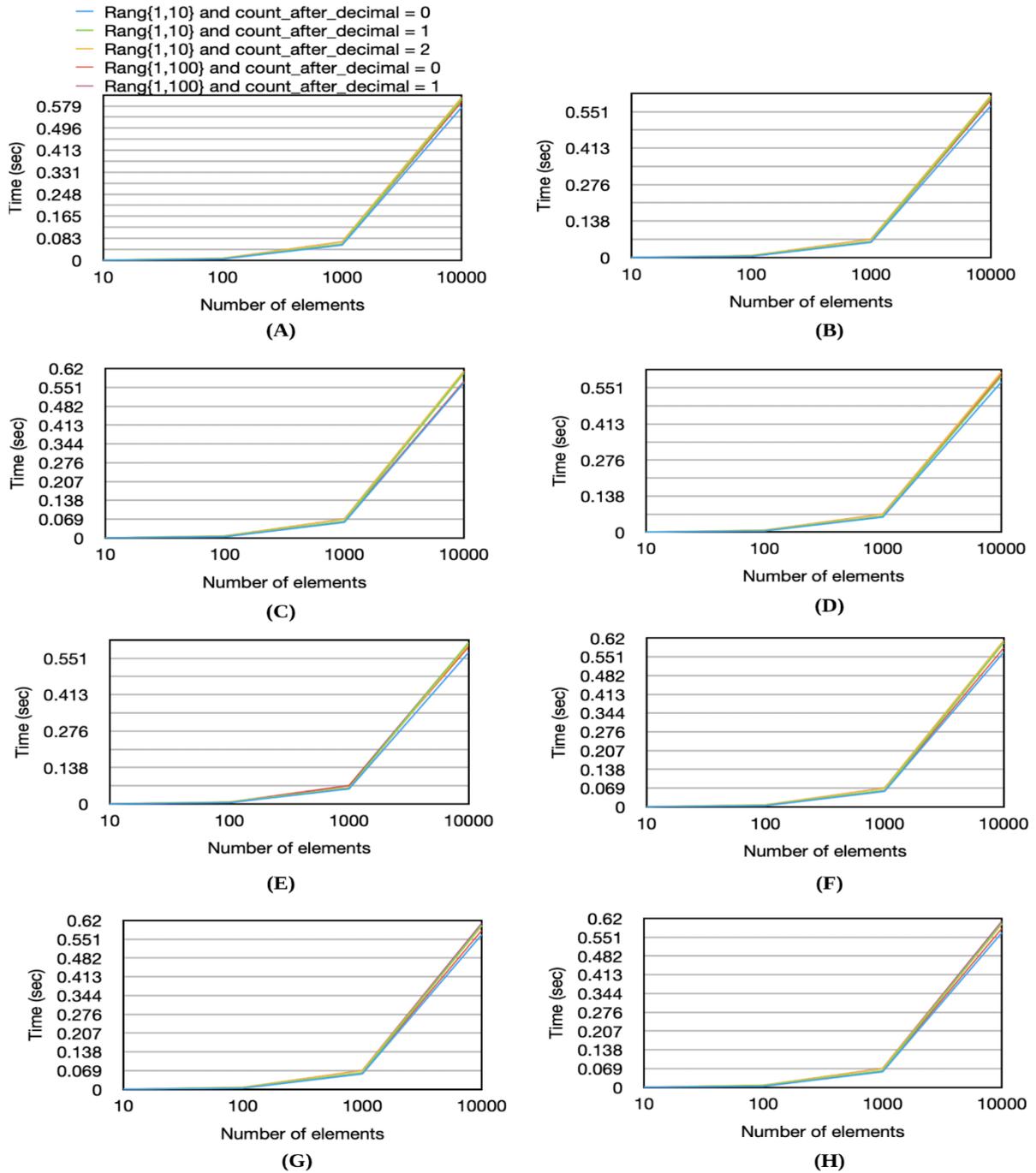

**Figure 5.** Graphs A-H represent the linear characteristics depicted by tables 4-11 respectively

**EXAMPLE 1**

As depicted in Figure 1, the array *arr* (defined above) is fed to the hashing cycle for sorting and the space *S* of 10x10 is initialized along with a vector H_Max of shape 10 and a space H_Min of shape 10x2. The further steps are as follows:

1. The first element of the array is '4.5', so:
   a. First, it will be multiplied by $10^1$ (as count after decimal is 1): $4.5 \times 10 = 45$
   b. Second, the number 45 will be converted to string using $F_{String}(45) = t = $ '45'.
   c. Third, we will increment the value in the memory block at row t[0] = 4 and column t[1] = 5 (at array index ( 4 , 5 ) ).
   d. Fourth, in the traverse map H_Max, as H_Max [ $F_{Numeric}$ (t[0])] < $F_{Numeric}$ (t[1]), then H_Max[ $F_{Numeric}$ (t[0])] will be set as $F_{Numeric}$(t[1]).
   e. Fifth, in traverse map H_Min, as H_Min [ $F_{Numeric}$ (t[0])][0] = = 0, then H_Min[ $F_{Numeric}$ (t[0])][1] will be set as $F_{Numeric}$(t[1]) and H_Min[$F_{Numeric}$( t[0] )][0] will be set to 1.

2. The first element of the array is '0.3', so:
   a. First, it will be multiplied by $10^1$ (as count after decimal is 1): $0.3 \times 10 = 03$
   b. Second, the number 03 will be converted to string using $F_{String}(03) = t = $ '03'.

c. Third, we will increment the value in the memory block at row t[0] = 0 and column t[1] = 3 (at array index ( 0 , 3 ) ).
d. Fourth, in the traverse map H_Max, as H_Max [ $F_{Numeric}$ (t[0])] < $F_{Numeric}$ (t[1]), then H_Max[ $F_{Numeric}$ (t[0])] will be set as $F_{Numeric}$(t[1]).
e. Fifth, in traverse map H_Min, as H_Min [ $F_{Numeric}$ (t[0])][0] = = 0, then H_Min[ $F_{Numeric}$ (t[0])][1] will be set as $F_{Numeric}$(t[1]) and H_Min[$F_{Numeric}$( t[0] )][0] will be set to 1.

3. The first element of the array is '2.3', so:
   a. First, it will be multiplied by $10^1$ (as count after decimal is 1): 2.3×10 = 23
   b. Second, the number 23 will be converted to string using $F_{String}$(23) = t = '23'.
   c. Third, we will increment the value in the memory block at row t[0] = 2 and column t[1] = 3 (at array index ( 2, 3 ) ).
   d. Fourth, in the traverse map H_Max, as H_Max [ $F_{Numeric}$ (t[0])] < $F_{Numeric}$ (t[1]), then H_Max[ $F_{Numeric}$ (t[0])] will be set as $F_{Numeric}$(t[1]).
   e. Fifth, in traverse map H_Min, as H_Min [ $F_{Numeric}$ (t[0])][0] = = 0, then H_Min[ $F_{Numeric}$ (t[0])][1] will be set as $F_{Numeric}$(t[1]) and H_Min[$F_{Numeric}$( t[0] )][0] will be set to 1.

4. The first element of the array is '8.8', so:
   a. First, it will be multiplied by $10^1$ (as count after decimal is 1): 8.8×10 = 88
   b. Second, the number 88 will be converted to string using $F_{String}$(88) = t = '88'.
   c. Third, we will increment the value in the memory block at row t[0] = 8 and column t[1] = 8 (at array index ( 8, 8 ) ).
   d. Fourth, in the traverse map H_Max, as H_Max [ $F_{Numeric}$ (t[0])] < $F_{Numeric}$ (t[1]), then H_Max[ $F_{Numeric}$ (t[0])] will be set as $F_{Numeric}$(t[1]).
   e. Fifth, in traverse map H_Min, as H_Min [ $F_{Numeric}$ (t[0])][0] = = 0, then H_Min[ $F_{Numeric}$ (t[0])][1] will be set as $F_{Numeric}$(t[1]) and H_Min[$F_{Numeric}$( t[0] )][0] will be set to 1.

5. The first element of the array is '7.0', so:
   a. First, it will be multiplied by $10^1$ (as count after decimal is 1): 7.0×10 = 70
   b. Second, the number 70 will be converted to string using $F_{String}$(70) = t = '70'.
   c. Third, we will increment the value in the memory block at row t[0] = 7 and column t[1] = 0 (at array index ( 7 , 0 ) ).
   d. Fourth, in the traverse map H_Max, as H_Max [ $F_{Numeric}$ (t[0])] <= $F_{Numeric}$ (t[1]), then H_Max[ $F_{Numeric}$ (t[0])] will be set as $F_{Numeric}$(t[1]).
   e. Fifth, in traverse map H_Min, as H_Min [ $F_{Numeric}$ (t[0])][0] = = 0, then H_Min[ $F_{Numeric}$ (t[0])][1] will be set as $F_{Numeric}$(t[1]) and H_Min[$F_{Numeric}$( t[0] )][0] will be set to 1.

6. The first element of the array is '9.2', so:
   a. First, it will be multiplied by $10^1$ (as count after decimal is 1): 9.2×10 = 92
   b. Second, the number 92 will be converted to string using $F_{String}$(92) = t = '92'.
   c. Third, we will increment the value in the memory block at row t[0] = 9 and column t[1] = 2 (at array index ( 9, 2 ) ).
   d. Fourth, in the traverse map H_Max, as H_Max [ $F_{Numeric}$ (t[0])] < $F_{Numeric}$ (t[1]), then H_Max[ $F_{Numeric}$ (t[0])] will be set as $F_{Numeric}$(t[1]).
   e. Fifth, in traverse map H_Min, as H_Min [ $F_{Numeric}$ (t[0])][0] = = 0, then H_Min[ $F_{Numeric}$ (t[0])][1] will be set as $F_{Numeric}$(t[1]) and H_Min[$F_{Numeric}$( t[0] )][0] will be set to 1.

7. The first element of the array is '4.5', so:
   a. First, it will be multiplied by $10^1$ (as count after decimal is 1): 4.5×10 = 45
   b. Second, the number 45 will be converted to string using $F_{String}$(45) = t = '45'.
   c. Third, we will increment the value in the memory block at row t[0] = 4 and column t[1] = 5 (at array index ( 4 , 5 ) ).
   d. This step will be skipped.
   e. This step will be skipped.

8. The first element of the array is '4.3', so:
   a. First, it will be multiplied by $10^1$ (as count after decimal is 1): 4.3×10 = 43
   b. Second, the number 43 will be converted to string using $F_{String}$(43) = t = '43'.
   c. Third, we will increment the value in the memory block at row t[0] = 4 and column t[1] = 3 (at array index ( 4 , 3 ) ).
   d. This step will be skipped.
   e. Fifth, in the traverse map H_Min, as H_Min [ $F_{Numeric}$ (t[0])][0] != 0 *and* H_Min [ $F_{Numeric}$ (t[0])][1] > $F_{Numeric}$ (t[1]) then H_Min[ $F_{Numeric}$ (t[0])][1] will be set as $F_{Numeric}$(t[1]).

9. The first element of the array is '8.0', so:
   a. First, it will be multiplied by $10^1$ (as count after decimal is 1): 8.0×10 = 80
   b. Second, the number 80 will be converted to string using $F_{String}$(80) = t = '80'.
   c. Third, we will increment the value in the memory block at row t[0] = 8 and column t[1] = 0 (at array index (8, 0) ).
   d. This step will be skipped.
   e. Fifth, in the traverse map H_Min, as H_Min [ $F_{Numeric}$ (t[0])][0] != 0 *and* H_Min [ $F_{Numeric}$ (t[0])][1] > $F_{Numeric}$ (t[1]) then H_Min[ $F_{Numeric}$ (t[0])][1] will be set as $F_{Numeric}$(t[1]).

10. The first element of the array is '3.2', so:
    a. First, it will be multiplied by $10^1$ (as count after decimal is 1): 3.2×10 = 32
    b. Second, the number 32 will be converted to string using $F_{String}(32)$ = t = '32'.
    c. Third, we will increment the value in the memory block at row t[0] = 3 and column t[1] = 2 (at array index ( 3, 2 ) ).
    d. Fourth, in the traverse map H_Max, as H_Max[ $F_{Numeric}$ (t[0])] < $F_{Numeric}$ (t[1]), then H_Max[ $F_{Numeric}$ (t[0])] will be set as $F_{Numeric}$(t[1]).
    e. Fifth, in traverse map H_Min, as H_Min[ $F_{Numeric}$ (t[0])][0] == 0, then H_Min[ $F_{Numeric}$ (t[0])][1] will be set as $F_{Numeric}$(t[1]) and H_Min[$F_{Numeric}$( t[0] )][0] will be set to 1.

The final result of this algorithm (Hashing Cycle) is given in Figure 2.